# Magnetic control of reflectionless states and an intrinsic-mode exceptional point in a magnon-phonon coupling system

Mingyuan Ma[1,#], Jiongjie Wang[2,#], Chong Chen[1], Weiqi Du[1], Liyang Liao[1], Yanzhang Cao[1], Peisen Liu[1], Boyuan Xiao[1], Deqi Tao[1], Shixuan Liang[1], Sulei Fu[1], Jiang Xiao[2,*], Cheng Song[1,*], Feng Pan[1,*]

[1] *Key Laboratory of Advanced Materials (MOE), School of Materials Science and Engineering, Tsinghua University, Beijing 100084, China*

[2] *Department of Physics and State Key Laboratory of Surface Physics, Fudan University, Shanghai 200433, China*

Reflectionless states (RLs) and exceptional points (EPs) are hallmark singular responses of non-Hermitian systems, with potential applications in sensing, wave control, and energy manipulation. Yet, realizing and independently tuning these states in a compact integrated platform remains challenging. Here, we demonstrate two-dimensional magnetic control of RLs and EPs in an on-chip magnon–phonon system based on a FeGaB/32°YX-$LiNbO_3$ heterostructure. The magnetic-field magnitude tunes the magnon–phonon detuning, driving a low-loss reflection zero across the real-frequency axis and producing two field-tunable RLs with near-vanishing reflection. A closed contour of the experimentally measured complex $S_{11}$ around the reflection-zero singularity accumulates a $-2\pi$ phase winding. Independently, rotating the magnetic field tunes the magnon-phonon coupling strength $g$, driving the intrinsic magnon–phonon eigenmode spectrum across an exceptional point, identified by the exchange between real-part and imaginary-part eigenvalue splitting. A unified one-port input–output model connects the measured complex $S_{11}$ response to the field-driven dynamics of the reflection zeros and the angle-tuned intrinsic magnon–phonon eigenmodes. These results establish a magnetically programmable non-Hermitian magnon–phonon platform for integrated wave control and compact vector magnetic sensing.

## 1. Introduction

Non-Hermitian systems, in which dissipation and environmental coupling render the spectral response complex, host a variety of unconventional wave phenomena, including bound states in the continuum [1,2], perfect absorption [3], topological braiding [4–6], reflectionless states [6,7], and exceptional points (EPs) [8,9]. The enigmatic acoustic silence in mountain valleys finds its quantum counterpart in non-Hermitian reflectionless states (RLs), where radiative and dissipative loss rates are balanced through critical coupling [10,11]. RLs correspond to scattering zeros that reach the real-frequency axis, producing vanishing reflection for a real incident wave. EPs are unique degeneracies at which multiple eigenvalues and eigenvectors coalesce [12,13], acting as critical points for phase transitions and enhanced sensing sensitivity. Observed across photonics [14,15], acoustics [16,17], magnetism [18], and electronics [19,20], RLs and EPs enable diverse applications in energy harvesting [21], sensing [22], and noise control [23]. However, their implementation is hindered by three fundamental limitations: (i) integration difficulty [24]; (ii) limited external tunability, including simultaneous tuning of the gain–loss contrast [25], the frequency detuning, and the coupling between different quanta [26]; and (iii) the need for complex excitation schemes based on waveform engineering, which are limited to low-quality-factor resonators in nano-photonics and other high-frequency systems [11]. Thus, realizing both RLs and EPs in a compact integrated platform with simple and continuous external tuning remains an important challenge.

Magnon–phonon coupling systems provide a promising route toward this goal. These systems are intrinsically non-Hermitian owing to the Gilbert damping of the magnetic layer and the loss of the acoustic cavity [27–29]. The alternating strain of a surface acoustic wave (SAW) cavity excites spin waves in the magnetic layer through the inverse piezoelectric and magnetostrictive effects—a process termed magnon–phonon coupling [30,31]. SAWs propagate on a solid chip at short acoustic wavelengths, providing a unique opportunity for miniaturized on-chip integration with a magnetic layer [32]. More importantly, an external magnetic field provides two complementary control degrees of freedom. Its magnitude tunes the magnon frequency through the Zeeman interaction, thereby controlling the magnon–phonon detuning, whereas its orientation modifies the magnetoelastic driving efficiency and hence the coupling strength $g$ [33–35]. This combination enables magnetic control over both the diagonal and off-diagonal terms of the effective coupled-mode problem without requiring optical pumping, active gain, or external phase-control networks. Recent work by Todaka et al. represents an important step in extending experimental studies of magnon-phonon coupling into the non-Hermitian regime [36]. By magnetically tuning the phase of a complex coupling constant, that work showed that its imaginary component provides an additional channel through which SAW phonons can interact with overdamped magnons even when their resonance frequencies are far apart. The result shows that strong magnetic damping can open, rather than merely suppress, a distinct interaction regime and establishes magnetoacoustic structures as a platform for exploring non-Hermitian hybrid dynamics.

Here, we construct a non-Hermitian magnon–phonon system by growing a $Fe_{70}Ga_{20}B_{10}$ (FeGaB) film on a SAW acoustic resonator, in which an external magnetic field provides two-dimensional regulation of the magnon frequency detuning (via the Zeeman shift) and the magnon–phonon coupling strength. On the one hand, at a critical magnetic-field magnitude, a near-vanishing reflection coefficient is achieved, as evidenced by $|S_{11}| < 2 \times 10^{-3}$ together with a $-2\pi$ phase winding of the measured complex $S_{11}$ around the reflection-zero singularity. On the other hand, rotating the magnetic field continuously tunes the magnon–phonon coupling strength and drives the intrinsic magnon–phonon eigenmodes across an exceptional point, as identified by the exchange between the real-part and imaginary-part eigenvalue splittings. A unified one-port input–output model connects the measured complex scattering response to these field-driven reflection-zero dynamics and the angle-controlled intrinsic eigenmodes. Our work demonstrates magnetic control over distinct non-Hermitian scattering and eigenmode singularities within a single magnon–phonon device, offering a route toward compact wave control and vector magnetic sensing.

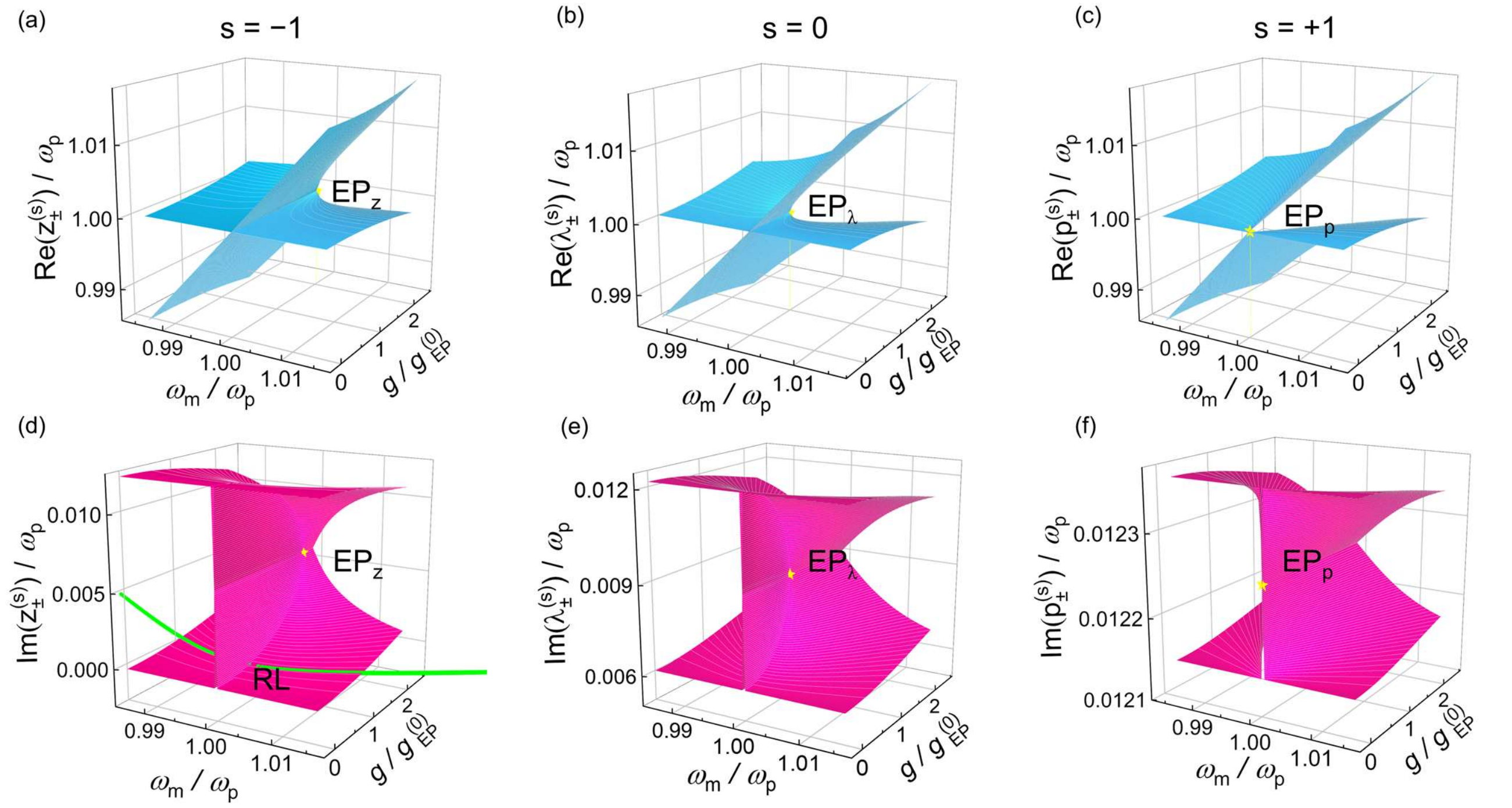

FIG. 1. Unified non-Hermitian spectral framework for the one-port magnon–phonon system. (a–c) Real and (d–f) imaginary parts of the complex-frequency spectra as functions of the normalized magnon frequency $\omega_m/\omega_p$ and coupling strength $g/g_{\mathrm{EP}}^{(0)}$. (a, d) Reflection-zero spectrum $z_{\pm}$, (b, e) intrinsic magnon–phonon eigenmode spectrum $\lambda_{\pm}$, and (c, f) scattering-pole spectrum $p_{\pm}$. Here, $g_{\mathrm{EP}}^{(0)} = |\kappa_m - \kappa_p|/2$ denotes the exceptional point threshold of the intrinsic magnon–phonon eigenmodes and is used as the common normalization for all panels. The marked yellow stars indicate the corresponding zero-, intrinsic-, and pole-spectrum exceptional point conditions. In (d), the green curve denotes the reflectionless state (RL) locus, where a reflection zero reaches the real-frequency axis, $\mathrm{Im}(z_{\pm}) = 0$.

## 2. Materials and methods

### 2.1 Theoretical framework of the one-port magnon–phonon coupling system

In contrast to conventional PT-symmetric systems [37], our platform operates in the intrinsically dissipative loss–loss regime, in which both the magnon and phonon modes carry intrinsic damping. This system exchanges energy with its environment through the electromagnetic waves used for the antenna [38,39]. To distinguish the different complex-frequency spectra involved in the one-port scattering problem, we introduce a unified effective matrix. The present magnon–phonon system can therefore be described theoretically by an input–output model with two complex-frequency modes, $\omega_p + i\kappa_p$ and $\omega_m +$

i$\kappa_m$, where $\omega_p$ ($\omega_m$) is the frequency of the phonon (magnon) mode and the imaginary part represents the respective intrinsic dissipation. The intrinsic phonon and magnon modes are coupled through a real-valued strength $g$, forming a 2 × 2 Hamiltonian [40].

$$H^{(s)} = \begin{bmatrix} \omega_p + i(\kappa_p + s\frac{\gamma}{2}) & g \\ g & \omega_m + i\kappa_m \end{bmatrix}, s = -1, 0, +1 \tag{1}$$

By employing the input-output formalism (Supplementary Information [41], S1), the $S_{11}$ parameters can be expressed as follows [42,43],

$$S_{11}(\omega) = -\frac{(\omega_\mathrm{m} + i\kappa_\mathrm{m} - \omega)\left(\omega_\mathrm{p} + i\kappa_\mathrm{p} - i\frac{\gamma}{2} - \omega\right) - g^2}{(\omega_\mathrm{m} + i\kappa_\mathrm{m} - \omega)\left(\omega_\mathrm{p} + i\kappa_\mathrm{p} + i\frac{\gamma}{2} - \omega\right) - g^2} \tag{2}$$

where $g$ is the magnon–phonon coupling strength, and $i\frac{\gamma}{2}$ labels damping introduced by detection channel. Here, $s$ is an index distinguishing three spectral problems rather than an experimentally tunable parameter: $s = -1$ describes the reflection-zero spectrum, $s = 0$ the intrinsic magnon–phonon eigenmode spectrum, and $s = +1$ the scattering-pole spectrum.

Fig. 1 compares these three complex spectra as functions of $\omega_m/\omega_p$ and $g/g_{\mathrm{EP}}^{(0)}$, where $g_{\mathrm{EP}}^{(0)} = |\kappa_m - \kappa_p|/2$. These contours collectively form a Riemann surface in the complex parameter space, where the paths of the magnon and phonon eigenfrequencies emerge as distinct spectral branches. EPs arise when both the real and imaginary parts coalesce at the spectral degeneracy point $\omega_m=\omega_p$, corresponding to the intersection of the two branches. External probing channel shifts the EP conditions of the reflection-zero and scattering-pole spectra away from that of the intrinsic eigenmodes as marked by the yellow stars in Fig. 1(a)-(c), whereas the intrinsic EP remains at $g/g_{\mathrm{EP}}^{(0)} = 1$ and $\omega_m/\omega_p = 1$. In addition, the green curve in Fig. 1(d) denotes the RL locus, defined by $\mathrm{Im}(z_\pm) = 0$, where a reflection zero reaches the real-frequency axis and the reflection vanishes. Therefore, the same magnon–phonon platform can host both reflectionless states and multiple classes of exceptional points, providing a unified setting for exploring non-Hermitian scattering and eigenmode singularities.

**2.2 Device fabrication and experimental configuration**

In the following experiments, FeGaB film was grown on an acoustic cavity (32°YX-$LiNbO_3/SiO_2/SiC$). A shear-horizontal (SH) SAW was excited by the interdigital transducers (IDTs). Fabrication details can be found in the Supplementary information S2. The resonant frequency of phonon is identified as: $\frac{\omega_p}{2\pi} = 4.1584$ GHz. Magnons were excited by the SH-SAW to form a coupled magnon-phonon system, as illustrated in Fig. 2(a). The SAW resonator exhibits a measured quality factor of approximately 81(Supplementary Information [41], S3), providing a finite acoustic dissipation channel required for the present non-Hermitian dynamics [44]. As the eigenfrequency of the system can be reflected by $S_{11}$ indirectly, we conducted microwave reflection $S_{11}$ measurements using a vector network analyzer (VNA), where external coupling was introduced. The magnetic field $H$ was swept from −500 to 500 Oe, and its orientation deviating from the wave vector $k$ was defined as $\theta$, as shown in Fig. 2(b).

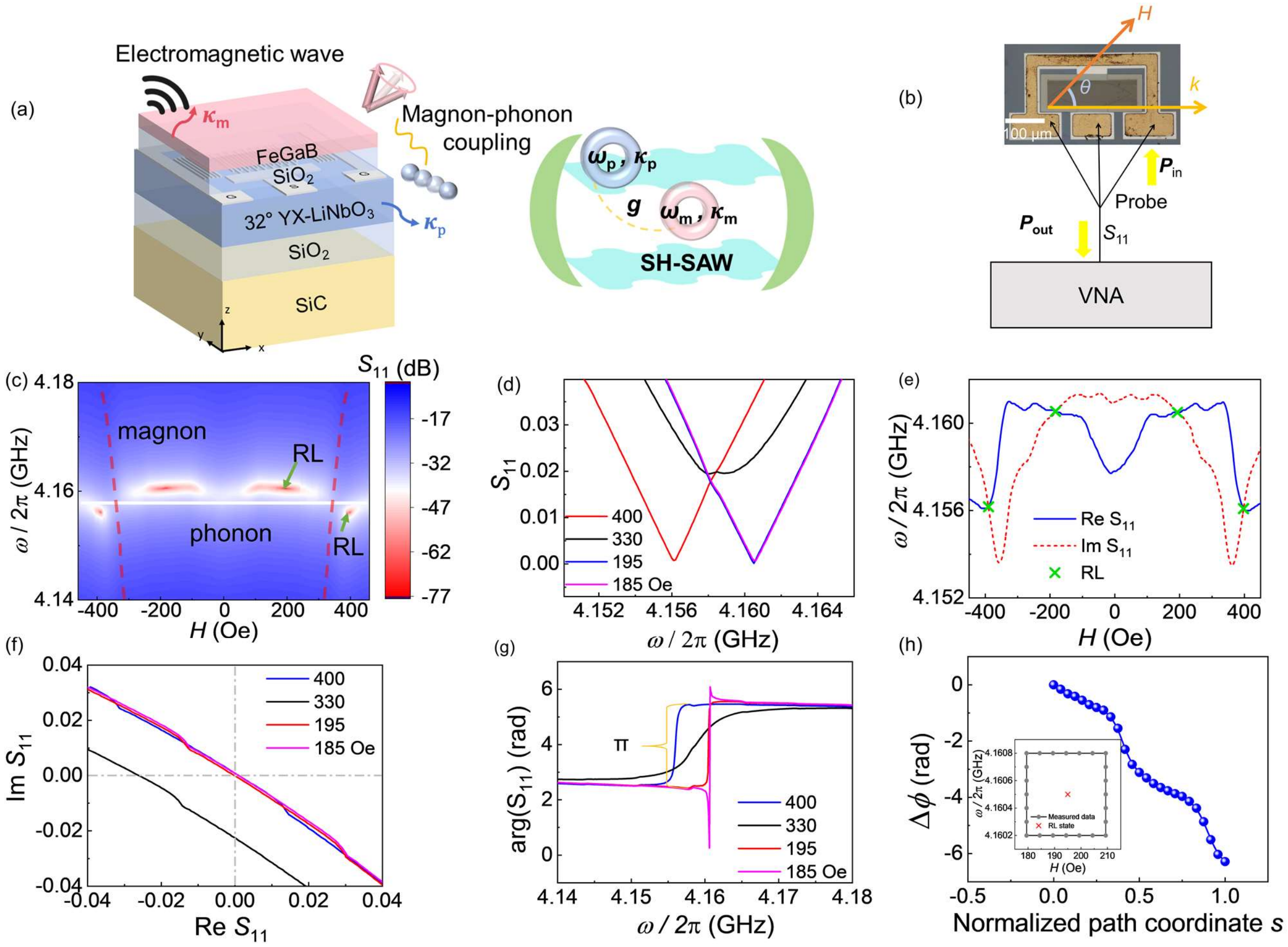

FIG. 2. Device configuration and experimental observation of magnetic-field-tuned reflectionless states. (a) Magnon-phonon coupling mechanism and schematic of the magnon-phonon coupling device (Supplementary Information, S2). (b) Experimental setup with $S_{11}$ measurements and optical image of the device, where $\theta$ is the angle between the magnetic field $H$ and wave vector $k$. (c) Experimental data mapping of $S_{11}$ as a function of magnetic field $H$ at $\theta = 0°$. The dashed line and solid line are the dispersion of magnon and phonon respectively. (d) Zoom-in view linear $S_{11}$ values at different $H$ when $\theta = 0°$. (e) Zero contours of $\mathrm{Re}(S_{11})$ and $\mathrm{Im}(S_{11})$ in the frequency–field plane; their intersections identify the reflection zeros reaching the real-frequency measurement plane. (f) Complex $S_{11}$ trajectories in the Argand plane near the RL, showing passage through the origin at the reflectionless condition. (g) Phase functions of frequency at different $H$ when θ = 0°. RLs exists at the magnetic field $H$ where a π phase jump occurs. (h) Accumulated phase along a closed contour enclosing the reflection zero in the frequency–field plane, exhibiting a $-2\pi$ phase winding corresponding to a winding number of $-1$. Inset is closed contour in the frequency–magnetic-field plane used for the phase-winding calculation, where gray circles denote the measured data points and the red cross marks the RL state.

## 3. Results and discussion

### 3.1 Magnetic-field-tuned reflectionless states

On the one hand, the frequency of magnons can be tuned by the magnetic field, enabling their coupling with phonons to produce spin wave resonance [45]. From the $S_{11}$ mapping with $H$ at $\theta$ = 0° in Fig. 2(c), it is interesting to note that there is a pronounced avoided crossing gap on the dispersion of phonons at $H$ = ±330 Oe with an $S_{11}$ dip of −22.8 dB. Notably, $S_{11}$ values adjacent to the gap at $H$ = 195 and 400 Oe are below −54 dB, indicating the existence of RLs. To further verify the RLs, a comparison of zoom in linear $S_{11}$ spectra at different $H$ values ($H$ = 185, 195, 330, and 400 Oe) are plotted in Fig. 2(d). Obviously, the linear $S_{11}$ values at $H$ = 195 and 400 Oe are below $2\times10^{-3}$, indicating near zero reflection coefficients for the RLs. We further examine the complex $S_{11}$ response to identify the underlying reflection zeros. Fig. 2(e) plots the zero contours of $\mathrm{Re}(S_{11})$ and $\mathrm{Im}(S_{11})$ in the frequency–field plane. Their intersections locate the conditions at which both components simultaneously approach zero, providing a direct experimental signature of a reflection zero reaching the real frequency measurement plane. Consistently, the corresponding trajectories of complex $S_{11}$ in the Argand plane (Fig. 2(f)) approach and pass through the origin near the RL condition. These complex-plane measurements establish a direct connection between the experimentally observed near-zero reflection and the reflection-zero spectrum introduced in Fig. 1.

In addition to the near-zero reflection coefficients, the RLs are accompanied by phase jumps. Surprisingly, the phase jump is almost equal to π at certain magnetic field, as illustrated in Fig. 2(g). Since the phase jump occurs at the saturated magnetic field of the magnetic layer, the phase change caused by the magnetic moment reversal owing to magnetic dynamics can be ruled out. The phases of $S_{11}$ at different $H$ values are also plotted in Fig. 2(g) for comparison. As the sweeping of $H$ from high field strength of 500 Oe, $\pi$-scale phase jumps appear when $H$ decreases to 400 Oe as depicted in Fig. 2(g). At this point, the reflection coefficient (<−70 dB) is nearly equal to zero, noted as RLs. As $H$ continues to decrease to 330 Oe where the spin wave resonance occurs, the phase changes gently with values smaller than π. The process of change in the intermediate magnetic field is omitted here. Interestingly, the process of phase changing has reversed as we continue lower down $H$. For example, a repeated π-scale phase jumps appear at $H$ = 195 Oe and a phase change > π appears at $H$ = 185 Oe. Further, along the clockwise contour enclosing the RL as illustrated in Fig. 2(h), the accumulated phase of the experimentally measured complex $S_{11}$ changes by $\Delta\phi = -2\pi$, corresponding to a winding number $v$ equals to $-1$. We found that two phase change processes here have one thing in common. The reflection coefficient is the lowest at the magnetic field where the π phase jump occurs, corresponding to the RLs. Then, we tune $\theta$ to determine whether RLs exist with changing $\theta$ (Supplementary Information, S4). It is found that the gap between the two RLs narrows with increasing $\theta$ and finally disappears.

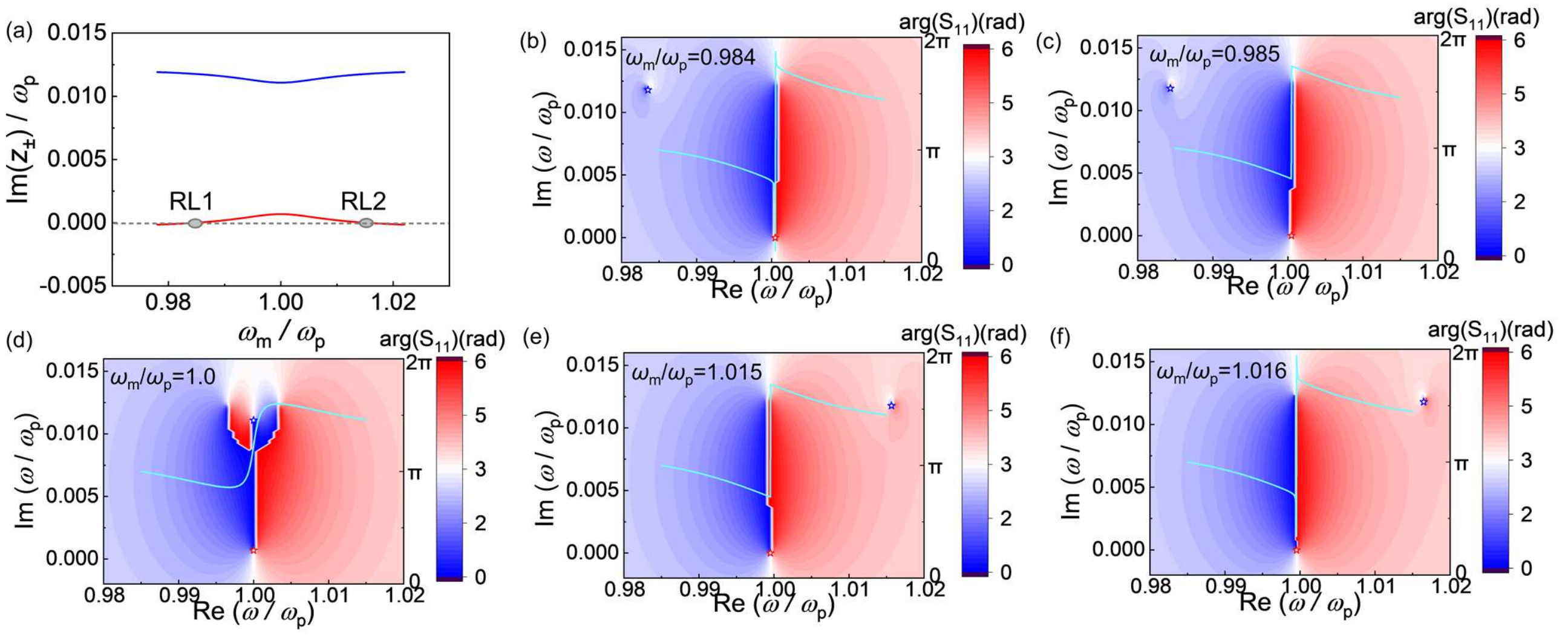

FIG. 3. Theoretical evolution of reflection zeros and the associated phase response during magnetic-field tuning. (a) Imaginary parts of the two reflection-zero branches as functions of the normalized magnon frequency $\omega_m/\omega_p$. The low-loss reflection zero crosses the real-frequency axis twice, producing RL1 and RL2, whereas the high-loss reflection zero remains away from the real axis. (b–f) Calculated phase distributions of $S_{11}$ in the complex-frequency plane and corresponding wrapped phase responses along the real-frequency axis at $\omega_m/\omega_p$ = 0.984, 0.985 ($RL_1$), 1.000 (magnon–phonon resonance), 1.015 ($RL_2$), and 1.016 respectively. The stars indicate the positions of the reflection zeros. The phase excursion

evolves from $> \pi$ to $\sim \pi$, then to $< \pi$, and reverses symmetrically across the second RL, illustrating the two real-axis crossings of the same low-loss reflection-zero trajectory.

To further clarify the origin of the field-tunable RLs, we calculate the evolution of the reflection zeros in the complex-frequency plane using the one-port input–output model (Supplementary Information[41], S5). As shown in Fig. 3(a), two reflection-zero branches exist owing to the coupled magnon–phonon dynamics. The low-loss reflection zero approaches and crosses the real-frequency axis twice as the normalized magnon frequency $\omega_m/\omega_p$ is tuned through magnetic field, whereas the high-loss reflection zero remains away from the real axis. The two crossings of the same low-loss reflection-zero trajectory correspond to $RL_1$ and $RL_2$, respectively. The associated phase response of $S_{11}$ is calculated at representative detuning, as shown in Fig. 3(b)–(f). At $\omega_m/\omega_p = 0.984$ ,where a low magnetic field $H$, the low-loss reflection zero lies slightly below the real-frequency axis and the phase excursion exceeds $\pi$, as shown in Fig. 3(b). As magnetic field strength increases to the first critical value, $\omega_m/\omega_p \simeq 0.985$, the zero reaches the real axis, giving $RL_1$ together with a phase excursion of approximately $\pi$ (Fig. 3(c)). With further increasing $H$, the zero moves above the real-frequency axis. At the magnon–phonon resonance, $\omega_m/\omega_p = 1$, the two reflection zeros have the same real part but remain separated in their imaginary parts, and the phase excursion decreases below $\pi$, as depicted in Fig. 3(d). This behavior indicates that the system does not reach a reflection-zero EP at this point.

Upon further increasing $H$, the same low-loss reflection-zero branch returns to the real-frequency axis at $\omega_m/\omega_p \simeq 1.015$, producing $RL_2$ with an approximately $\pi$ phase excursion (Fig. 3(e)). Beyond $RL_2$, for example at $\omega_m/\omega_p = 1.016$, the zero again moves below the real-frequency axis and the phase excursion recovers to a value larger than $\pi$ (Fig. 3(f)). This evolution demonstrates that the experimentally observed double-RL structure originates from two real-axis crossings of a single low-loss reflection-zero trajectory as the magnon frequency is swept across the phonon resonance, rather than from two independent hybrid branches separately satisfying the RL condition. The calculated phase evolution is also consistent with the experimentally observed restructuring of the $S_{11}$ phase in Fig. 2(g).

### 3.2 Magnetic-field-orientation-tuned exceptional point

To further validate what happens at the gap in Fig. 2(c), the $S_{11}$ lines of $H = \pm 330$ and $-500$ Oe are extracted from Fig. 2(c) and plotted in Fig. 4(a). It is found that the phonon frequency splits at $H = \pm 330$ Oe, but no splitting and a single Lorentzian line shape are maintained at 500 Oe, indicating a magnon-phonon coupling with $g$ larger than dissipation rates difference at the resonant field of the spin wave. On the other hand, the emergence of EPs is strongly dependent on the magnon-phonon coupling strength $g$. As a control experiment, the angle $\theta$ was adjusted to modify $g$ (Supplementary Information[41], S6), which is sensitive to the magnitude of the magnetoelastic driving field $h$, $g \propto h$. The $h$ value is determined by the magnetization orientation and the dominant strain component $\varepsilon_{xy}$ associated with the SH-SAW [46]. $\theta$ is changed from −20° to 20° in a 5° step to tune $g$ for EPs.

The evolution of the magnon-phonon spectral features under angular modulation is shown in Fig. 4(b), where the $S_{11}$ spectra at positive spin-wave resonance fields are plotted across $\theta = -20°$ to 20°. The angular-driven transition is vividly captured in Fig. 4(b), where the $S_{11}$ spectra at the magnon resonance fields shows a pronounced angle-dependent evolution of the hybrid spectral response. We attributed the strong in-plane anisotropy to the non-uniform magnetic film growing on the IDT. As $\theta$ changes from −20° to 20°, we can find that the $S_{11}$ frequency splits with two resonance peaks when $|\theta| \leq 10°$ and merges with a single Lorentzian peak when $|\theta| > 10°$. The merging of the measured resonance features provides an experimental signature of the changing coupling regime. To quantitatively describe the experimental response, the complex $S_{11}$ spectra are fitted using the one-port input–output model introduced above (Supplementary Information [41], S7). Using the extracted parameters, the calculated $S_{11}$ mapping at $\theta = 0°$is shown in Fig. 4(c). The model reproduces the magnon–phonon anticrossing as well as the field-dependent reflection features observed experimentally, supporting the use of the same fitted parameters for reconstructing the complex-frequency spectra. The fitted coupling strength $g$ and the three corresponding EP thresholds are summarized as functions of $\theta$ in Fig. 4(d). Here, $g_{\mathrm{EP}}^{(-1)}$, $g_{\mathrm{EP}}^{(0)}$, and $g_{\mathrm{EP}}^{(+1)}$correspond to the reflection-zero, intrinsic-eigenmode, and scattering-pole spectra, respectively. Over the measured angular range, $g$ remains below the reflection-zero EP threshold and above the scattering-pole EP threshold, whereas it crosses the intrinsic-mode threshold $g_{\mathrm{EP}}^{(0)}$ between 10°and 15°. This comparison therefore identifies the experimentally accessed degeneracy as an EP of the intrinsic magnon–phonon eigenmode spectrum. The corresponding complex intrinsic eigenvalues $\lambda_{\pm}$, reconstructed from the same fitted parameters, are plotted in Fig. 4(e) and (f). At small $|\theta|$, where $g > g_{\mathrm{EP}}^{(0)}$, the real parts of $\lambda_{\pm}$ are split while their imaginary parts are nearly coalesced, corresponding to the strong-coupling side of the non-Hermitian transition. As $|\theta|$ increases and $g$ decreases, the real-part splitting gradually closes. After crossing the EP between 10° and 15°, the real parts coalesce whereas the imaginary parts separate. The exchange from real-part splitting to imaginary-part splitting constitutes the characteristic complex eigenvalue signature of the exceptional point. Therefore, the angle dependent $S_{11}$ measurements, the extracted coupling strength relative to the three candidate EP thresholds, and the reconstructed complex eigenvalues consistently demonstrate a magnetic-field-orientation-driven intrinsic magnon–phonon EP.

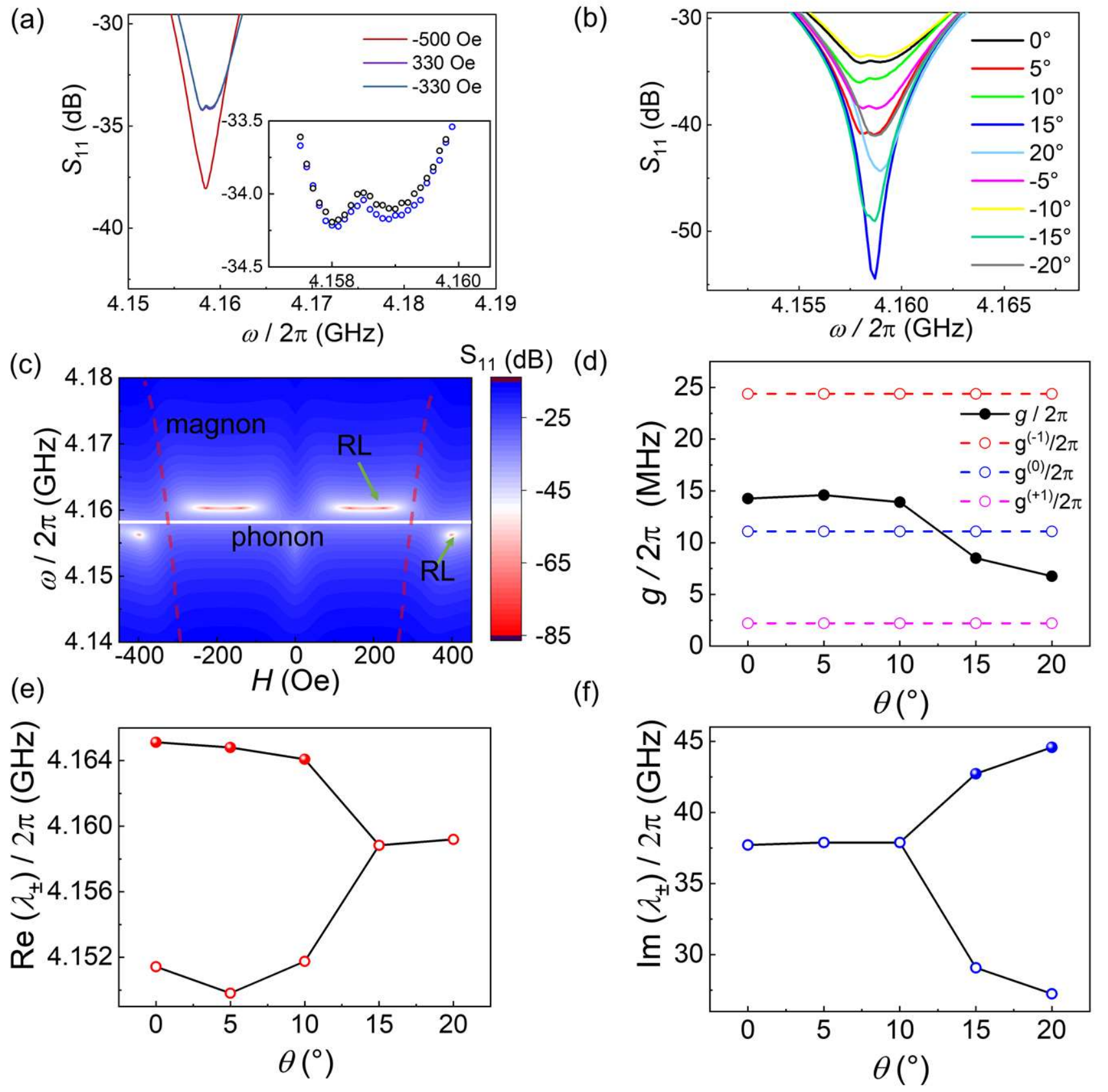

FIG. 4. Magnetic-field-orientation-tuned exceptional point in the magnon–phonon coupling system. (a) Measured $S_{11}$ spectra at the magnon–phonon resonant fields and at an off-resonant magnetic field. (b) Measured $S_{11}$ spectra at the magnon–phonon resonance for different in-plane magnetic-field orientations $\theta$. (c) Calculated $S_{11}$ mapping as a function of frequency and magnetic field at $\theta = 0°$, obtained from the one-port input–output model using the extracted system parameters. The dashed and solid lines denote the uncoupled magnon and phonon dispersions, respectively. (d) Fitted magnon–phonon coupling strength $g$ and the EP thresholds of the reflection-zero, intrinsic-eigenmode, and scattering-pole spectra as functions of $\theta$. (e, f) Real and imaginary parts of the reconstructed intrinsic magnon–phonon eigenvalues $\lambda_{\pm}$ as functions of $\theta$, respectively.

## 4. Conclusion

In summary, we experimentally reveal reflectionless states and an exceptional-point transition in a dissipative magnon–phonon coupling system. The RLs originate from the magnetic-field-driven motion of a low-loss reflection zero across the real-frequency axis, producing two tunable zero-reflection conditions accompanied by a nontrivial phase winding. By rotating the magnetic field, the magnon–phonon coupling strength is continuously varied, leading to the coalescence and subsequent exchange of the real and imaginary parts of the intrinsic complex eigenfrequencies across an EP. These findings uncover the non-Hermitian scattering and eigenmode dynamics hidden in magnon–phonon hybridization and provide a compact solid-state route for controlling singular wave responses. The strong dependence of RLs on field magnitude and of the EP on field orientation also suggests potential for integrated magnetic sensing and vector-field detection.

*Acknowledgement*—Mingyuan Ma and Jiongjie Wang contributed to this work equally. This work was supported by National Key R&D Program of China (Grant No. 2024YFA1408503), National Natural Science Foundation of China (Grant No. 52225106, 12241404, 52561160112, and T2394471), Tsinghua University Initiative Scientific Research Program, Natural Science Foundation of Beijing Municipality (Grant No. F251037), and the Open Fund of the State Key Laboratory of Spintronics Devices and Technologies (Grants No. SPL-2401). We acknowledge TuoTuo technology for lithography technical support.

[#]Mingyuan Ma and Jiongjie Wang contributed to this work equally.

[*]Corresponding author: xiaojiang@fudan.edu.cn; songcheng@mail.tsinghua.edu.cn; panf@mail.tsinghua.edu.cn